\documentclass[runningheads]{llncs}
\usepackage{amsfonts}
\usepackage{bm}
\usepackage{multirow}
\usepackage{graphicx}
\begin{document}
\title{Heard More Than Heard: An Audio Steganography Method Based on GAN}
\author{Dengpan Ye \and
Shunzhi Jiang\and
Jiaqin Huang }
\authorrunning{Denpan Ye \and Shunzhi Jiang et al.}
\institute{Key Laboratory of Aerospace Information Security and Trusted Computing, Ministry of Education, School of Cyber Science and Engineering, Wuhan University, Wuhan, China
\email{\{dpye,jsz\}@whu.edu.cn}}
\maketitle              % typeset the header of the contribution
\begin{abstract}
Audio steganography is a collection of techniques for concealing the existence of information by embedding it within a non-secret audio, which is referred to as carrier. Distinct from cryptography, the steganography put emphasis on the hiding of the secret existence. The existing audio steganography methods mainly depend on human handcraft, while we proposed an audio steganography algorithm which automatically generated from adversarial training. The method consists of three neural networks: encoder which embeds the secret message in the carrier, decoder which extracts the message, and discriminator which determine the carriers contain secret messages. All the networks are simultaneously trained to create embedding, extracting and discriminating process. The system is trained with different training settings on two datasets. Competed the majority of audio steganographic schemes, the proposed scheme could produce high fidelity steganographic audio which contains secret audio. Besides, the additional experiments verify the robustness and security of our algorithm.
\end{abstract}
\section{Introduction}
Steganography is the science to conceal secret messages in the carriers through slightly modifying the values which hard to detect by human perception. Similar to cryptography, the steganography provides methods for secret communication. Distinct from the cryptography method which focuses on the \textit{authenticity} and \textit{integrity} of the messages, steganography aims to \textit{hide the existence of the secret}. Massive surveillance operations have shown that the mere existence of meta-data communication could lead to privacy leakages even if the content is unknown. Therefore, steganography is necessary for private communication.\par
Audio steganography could be used in the watermark, copyright protection and secret transmission and many other applications. In general, the sender uses a steganographic algorithm to conceals a secret message into carrier audio which sounds unaltered to external detectors. The main effort in audio steganography is to minimize the perturbations within carrier audio when the secret is embedded within, while allowing for recovery of the secret message. Then the audio with the secret message was transmitted in public channel. The receiver intercepts the audio and extracts the secret message with the decoding algorithm and an established shared key.\par
Most existing audio steganography methods could be divided into three categories: temporal domain, transform domain and wavelet domain\cite{Audio ste}. In the temporal domain, most common methods encode message in the Least Significant Bit(LSB)\cite{LSB} of individual sound samples with the equivalent secret message binary sequence. In the transform domain, the methods use the masking effect of human auditory system and make the weak frequencies near the strong resonance frequencies inaudible, which including tone insertion, spread spectrum, phase coding\cite{TD2}. In the wavelet domain, the methods use discrete wavelet transform which could decompose the signal into high frequency and low frequency parts. The secret information can be embedded into discrete wavelet coefficients by combining with wavelet energy, masking effect, adaptive LSB\cite{WS}.\par
Distinct from the most audio steganography methods which mainly depend on human handcraft, we introduce the idea of automatically generated steganography algorithm based on deep learning. In this work, we try to obtain the audio steganographic algorithm by training generative adversarial networks(GANs)\cite{GAN}, which have proved to be competitive models on synthesis tasks in recent researches. The proposed scheme aims to conceal the secret audio within carrier audio. Thus, the task of the training is discriminative, the encoder takes in carrier audio and secret message and produces steganographic audio, while the discriminator tries to learn the weakness of generator, resulting in the ability to determine whether the audio contains secret. The model consists of three modules: the encoder module is used for synthesizing steganographic audio, while the discriminator detects the synthesis audio, the decoder is used for extracting the secret message. \par
In this paper, we proposed an audio steganography scheme which produces the steganographic audio through a novel GAN model and without human handcraft. We show that our scheme could successfully work in practice, the encoder could produce steganographic audio which contains secret audio, the decoder could decode the steganographic audio to a carrier and the less distortion secret audio, which means heard more than heard. The rest of the paper is organized as follows. Section 2 discusses the theory of steganography and GANs. Section 3 describes the proposed steganography scheme. In Section 4, we show the experiment results and discussions. Finally, conclusions and future works are given in Section 5.\par
\section{Related work}
\subsection{Steganagraphy and Steganalysis}
Steganography is the science of covered or hidden writing. The goal of steganography is covertly communicate secret messages. As shown in Figure. 1, there are three parts in this communicate: sender, receiver and external detector. To escape from the detection of external detectors, the main effort in steganography is to minimizing the perturbations between steganographic data and original carrier.
\begin{figure}
	\begin{center}
	\centering
	\includegraphics[width=0.9\linewidth]{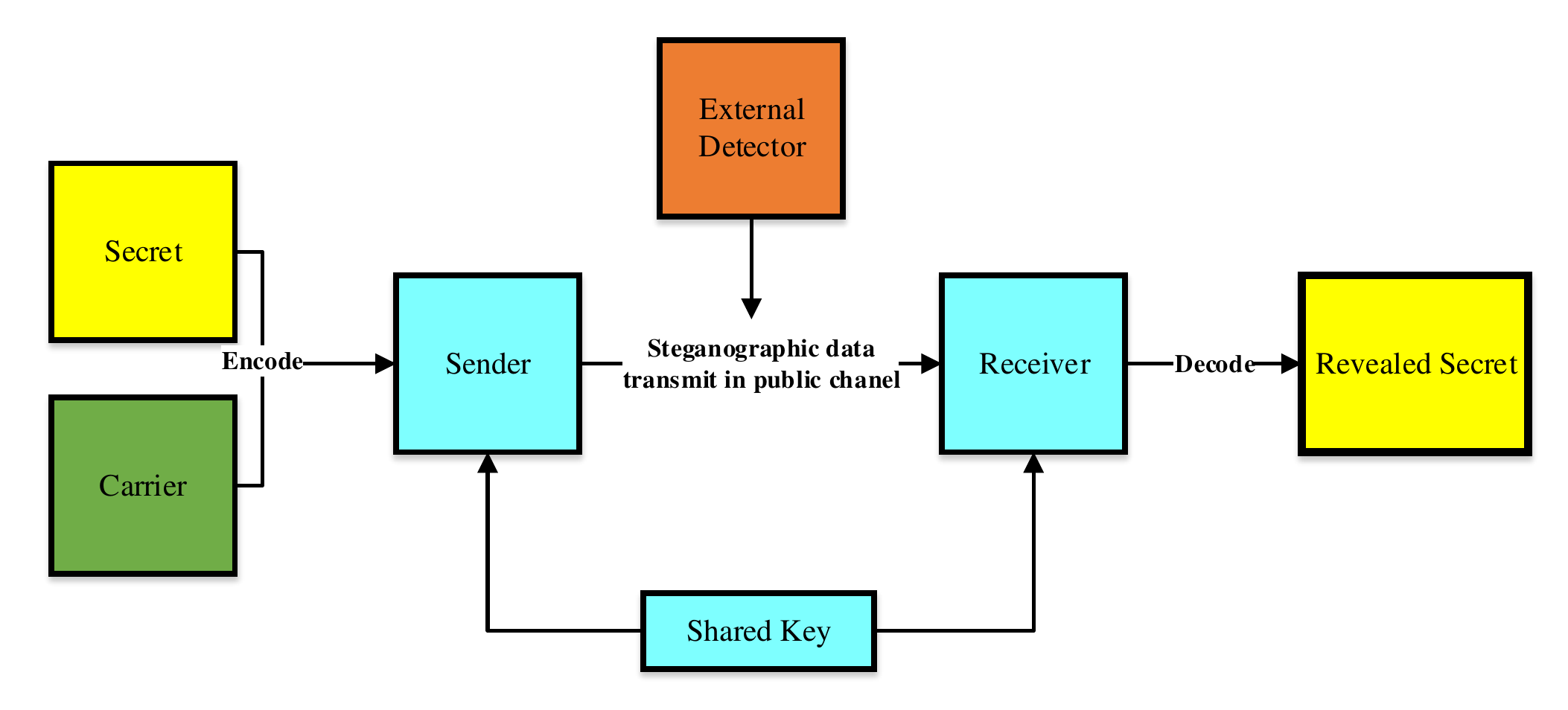}
	\caption{The three parts in this communicate: sender, receiver and external detector. The sender uses a steganographic algorithm to conceal a secret message into carrier which unaltered to external detectors. The receiver intercepts the data and extracts the secret message with the decoding algorithm and an established shared key.} \label{fig1}
	\end{center}
\end{figure}
There are many works in steganography, where most of them manipulate LSB algorithm to embed messages on gray and color images. Then some advanced method emerged, HUGO\cite{HUGO} preserves the input statistics by assigning costs to pixels to generate better steganography functions. WOW\cite{WOW} uses a bank of directional filters to estimate the distortion of predictable regions and embeds messages according to regions of complexity, the more texturally complex a region is, the more pixel values within that region will be modified. S-UNIWARD\cite{S-UNIWARD} uses a universal distortion function to avoiding the smooth of images. Similar to WOW, S-UNIWARD minimizes this distortion function, and embed messages in noisy regions or complex textures.\par
Corresponding to steganography, steganalysis takes the role of the external detector which detects hidden data. Usually, this task is formulated as a binary classification problem to distinguish between the carrier and steganographic data. Spatial Rich Model(SRM)\cite{srm} are the most famous steganalyzers which constructed by assembling a rich model as a union of many diverse submodels formed by joint distributions of neighboring samples from quantized noise residuals obtained using linear and non-linear high-pass filters\cite{holub}. The diversity of filters make the success of the so-call rich media models. For example, the two kernels:\par
\begin{equation}
K_3 = {\frac{1}{4}\left(\begin{array}{ccc}
	-1 & 2 & -1\\
	2 & -4 & 2\\
	1 & 2 & -1
	\end{array}
	\right)}
\end{equation}
\begin{equation}
K_5 = {\frac{1}{12}\left(\begin{array}{ccccc}
	-1 & 2 & -2 & 2 & -1\\
	2 & -6 & 8 & -6 & 2\\
	-2 & 8 & -12 & 8 & -2\\
	2 & -6 & 8 & -6 & 2\\
	-1 & 2 & -2 & 2 & -1
	\end{array}
	\right)}
\end{equation}
respectively, $K_3$ and $K_5$ predict the value of the central pixel from its local $3\times3$ and $5\times5$ neighborhoods. The unions of different filter residuals in rich model also wildly used in deep learning based steganalysis schemes\cite{qiannet,yenet,jiangnet}. In this work, to obtain better steganographic and security performance, we use stegnalyzer as discriminator in Section 3.2.
\subsection{Generative adversarial networks}
Generative adversarial networks(GANs) have proved to be competitive models on synthesis tasks in recent research. Typically, a GAN contains two parts: generator and discriminator. The generator captures the training data distribution and learns to create fake data different from the training data, while the discriminator learns to determine whether the input data is real or fake. The generator and discriminator are simultaneously trained via an adversarial process until the generator could produce high-quality fake data. The generator $G$ and discriminator $D$ play the following two-player minmax game with value function $V(G, D)$:
\begin{equation}
\min\limits_{G}\max\limits_{D}V(G,D) = \mathbb{E}_{\bm{x} \sim  p_{data(\bm{x})}}[\log{D(\bm{x})}] + \mathbb{E}_{\bm{z} \sim  p_{\bm{z}(\bm{z})}}[\log{1-D(G(\bm{z}))}]
\end{equation}
In the actual model training process, the equation may not provide sufficient gradient for $G$ to learn well. The capability of $G$ is poor, $D$ could reject samples with high confidence. Therefore, training $G$ to maximize $\log{D(G(\bm{z}))}$ could obtain better effect. In this work, we use an encoder network as G which produce steganographic audio. Besides, we use a steganalyzer as D to form the adversarial process. In addition, we use a decoder network which decodes and extracts the secret message in the steganographic audio. All the networks are simultaneously trained until the G could produce high fidelity data, the decoder could obtain less distorted secret message and D distinguish carrier and steganographic audio with high confidence. The trained steganalyzer(D) in GAN could use as a monitor on the listener.\par
\section{Audio Steganography scheme based on GAN}
The architecture of our Steganography scheme is shown in Figure. 2. In this work, our model is composed of three parts: encoder decoder and steganalyzer. Autoencoder networks are closely model to cryptography and steganography. Therefore, we use two networks which correspond conceal and reveal process in steganography. To obtain better steganographic and security performance, we use a steganalyzer as discriminator in the adversarial training process.\par
\begin{figure}
	\includegraphics[width=0.9\linewidth]{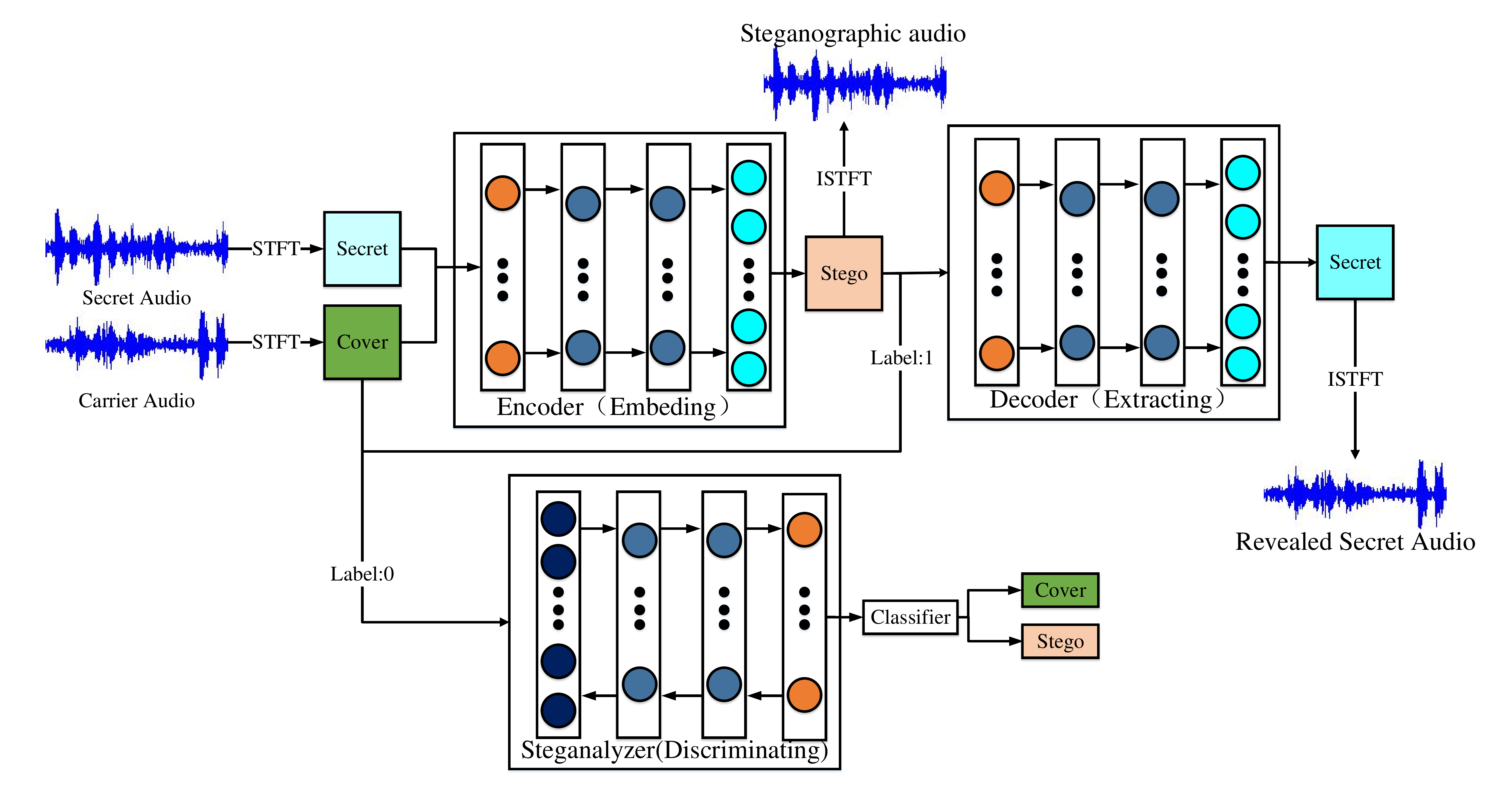}
	\centering
	\begin{center}
	\caption{The three parts in our scheme: encoder decoder and steganalyzer. The encoder accepts carrier audio and produces steganographic audio. The decoder extracts the secret message and produces a revealed secret audio. A CNN based steganalyzer is used as the discriminator of our GAN steganography model. All the networks are simultaneously trained to create embedding, extracting and discriminating process.} \label{fig2}
	\end{center}
\end{figure}
\subsection{Encoder and Decoder networks}
The encoder accepts carrier audio and secret audio as input. We apply the Short Time Fourier Transform(STFT) to transform the audio time domain to the frequency domain and get two spectrograms. The carrier audio spectrogram is concatenated with secret audio spectrogram in the first layer of the encoder. The encoder produces a steganographic spectrogram and passing this to the decoder. The decoder network produces a revealed secret spectrogram and then transform to secret audio by using Inverse Short Time Fourier Transform (ISTFT). Note that, the spectrogram produced by the decoder is the same size as the spectrogram which takes in by encoder.\par
The convolutional layers\cite{CNN} and inception modules are introduced to the encoder. The inception model\cite{Inception} are widely used in classification tasks. Such a structure contains several convolution kernels can fuse feature maps with different receptive field sizes. To accelerate the training process, we add batch normalization\cite{BN} layers in our model. The detail structure of the encoder and decoder network are described in Table. 1. The encoder and decoder networks form the generator of our GAN steganography model.\par
\subsection{Steganalyzer}
The security is the crucial point in steganography scheme. The steganography method like HUGO slightly alters the least significant bits of the data which hardly detects by human perception. However, the existing steganalysis could detect it well by using machine learning or deep learning model. In our work, we use a CNN based steganalyzer as the discriminator of our GAN steganography model. The input of steganalyzer composed of two parts: the spectrogram produced by encoder which labeled 1 and the carrier spectrogram which labeled 0. The output of steganalyzer is a confidence score of how likely this is a carrier or steganographic spectrogram. The detail structure of the steganalysis network is described in Table. 1. We employ a shallow but wide framework with some convolution layers and wide fully connected layers to detect the loss of tiny residual produce by embedding. To verify the security, we use different steganalyzer to detect our steganography method in Section 4.\par
\begin{table}[]
	\centering
	\caption{ The detail structure of the encoder decoder and steganalysis network. The Convblock-n represents n$*$3$*$3conv+BN+Relu. The InceptionBlock-n represents n$*$3$*$3Conv+BN+Relu. The HPF represents the High Pass Fliter in SRM. The FC represents the fullyconected layer. }
	\begin{tabular}{ccccc}
	\hline
		layers & Encoder & Decoder & Stegalyzer   \\
		\hline
		1      & input              & input              & input         \\
		2      & STFT               & Convblock-16       & HPF/Random        \\
		3      & Convblock-16       & Convblock-32       & Convblock-16        \\
		4      & InceptionBlock-32  & Convblock-64       & Convblock-32       \\
		5      & InceptionBlock-64  & Convblock-128      & Convblock-64       \\
		6      & InceptionBlock-128 & Convblock-64       & FC       \\
		7      & InceptionBlock-64  & Convblock-32       & FC       \\
		8      & InceptionBlock-32  & Convblock-1        & FC       \\
		9      & Convblock-32       & ISTFT              & Softmax        \\
		10     & Convblock-16      & Output             &   Output       \\
		11     & Convblock-16&    &          \\
		12     & Convblock-1&   &          \\
		13     & Output&        &          \\
			\hline
	\end{tabular}
\end{table}
\subsection{Training process and Loss function}
At the beginning of training, a human can easily separate carrier audio from steganographic audio, as the encoder has not learned the ability of embedding. The steganalyzer is like the discriminator in GAN, where we tie its predictive ability to embedding capacity of the encoder by labeled the carrier and steganographic spectrogram. As the train continues, the steganalyzer become stronger, and then the weights of encoder updates the parameters based on the loss of steganalyzer.\par
Denoting $\theta_{E},\theta_{D},\theta_{S}$ as the parameters of encoder decoder and steganalyzer. Let $A_c$, $A_i$, $A_s$, $A_r$ for carrier audio, secrect audio, steganographic audio and reveal secrect audio. Let $O_e(\theta_{E}, A_c, A_i)$, $O_d(\theta_{D}, A_s)$, $O_s(\theta_{S}, A_c, A_s)$ as the output of encoder, decoder and steganalyzer. We have:
\begin{equation}
O_d(\theta_{D}, A_s) = O_d(\theta_{D}, O_e(\theta_{E}, A_c, A_i))
\end{equation}
\begin{equation}
O_s(\theta_{S}, A_c, A_s) = O_s(\theta_{S}, A_c, O_e(\theta_{E}, A_c, A_i))
\end{equation}
Let $L_e$, $L_d$, $L_s$ denote the loss of encoder, decoder and steganalyzer. Common to the discriminator in GAN, we set steganalyzer loss to be cross entropy loss. The Euclidean distance $d$ between $A_i$ and $A_r$ is used in decoder reveal loss. The encoder loss which is vital in our steganography scheme is given by the sum of encoder loss and steganalyzer loss. $\lambda_a$, $\lambda_b$, $\lambda_c$ represent the weight given to each respective loss. Then we have:
\begin{equation}
L_s(\theta_{S}, A_c, A_s) = -y\cdot log(O_e(\theta_{E}, x))-(1-y)\cdot log(1-O_e(\theta_{E}, x))
\end{equation}
\begin{equation}
L_d(\theta_{E}, \theta_{S}, A_c, A_s) = d(A_r, A_i)
\end{equation}
\begin{equation}
L_e(\theta_{E}, A_c, A_i) = \lambda_a\cdot d(A_c, A_s) + \lambda_b\cdot L_s+ \lambda_c\cdot L_d
\end{equation}
After the model training is completed, the sender uses the encoder to embed a secret audio and send it to the receiver with some confused samples in the public channel. The receiver uses the decoder to decode the message. The steganalyzer could be used as a monitor at the side of the receiver cause its well discriminating performance under the training of GAN. The three parts in our steganography are like the shared key in cryptography which should perform offline. The cost of sending the model information is low, with an average of 10MB.
\section{Experiments}
In this section, extensive experiments are carried out to
demonstrate the effectiveness of our method. We implemented our scheme with different training settings under two audio dataset: TIMIT and LibriSpeech. The code and experimental data are available on our github.
\subsection{Dataset}
We implemented our work under TIMIT\cite{TIMIT} and LibriSpeech\cite{Lri} dataset. The TIMIT dataset contains broadband recordings of 630 speakers of eight major dialects of American English, each reading ten phonetically rich sentences. LibriSpeech contains approximately 1000 hours of reading English speech. Due to the TIMIT datasets have different audio lengths, the audio lengths with less than 32640 are omitted. The audio lengths larger than 32640 were segmented, then the audios were aggregated together to form the new TIMIT dataset. The audio length of LibriSpecch was uniform and larger than 32640, a new equal-length dataset was formed by selecting the first 32640 bits for each audio. For both the TIMIT and LribriSpeech datasets, we used 10,000 audio samples and split in quarter, creating a training set and a test set.  
\subsection{Implementation Details}
We implemented the code using PyTorch\cite{pytorch}, on a workstation with a 2080Ti GPU. The input of encoder was a design pair of audio which corresponding to carrier and secret. The secret audios were random chose in the training set so that the encoder does not learn a specific function associated with a specific group of information. Each audio was sampled at 22kHz and represented as its spectrum by STFT with 512 FFT frequency bins and 10ms sliding window. The audio after STFT was complex and in the form of $a + bi$. In order to normalize the training, it needs to be expressed as real numbers. There were two strategies to spilt $a + bi$: the real part and imaginary part, or amplitude and phase part. Both schemes divide a complex matrix into two real matrix representations. Considering that each channel in the image processing has the same status. Besides, the magnitude and phase have different dimensions which can not be directly overlapped and processed at the same time, the first strategy was chosen. The complex number was divided into a real part and imaginary part: extract $a$ and $b$, then overlapped them to produced a two-channel matrix. In the inverse transformation process of exacting, the two matrices are merged into a complex number matrix and decoded. We balanced between the carrier and steganographic losses and using $\lambda_a$ =0.6, $\lambda_b$ = 0.8, $\lambda_b$ = 1 for TIMIT and $\lambda_a$ =0.8, $\lambda_b$ = $\lambda_c$ = 1 for LibriSpeech. All models were trained using Adam for 75 epochs and 150 epochs with a learning rate of 0.0001.\par
\subsection{Performance Experiments}
We used MSE(Mean Square Error) and SNR(Signal to Noise Ration) as metrics to measure our model’s performance. MSE was used to measure the audio loss while SNR was used to measure the quality of the audio. Two settings are used during the process of training: RAN which meant the first layer of our steganalyzer in GAN was randomly initialized; HPF which means the first layer is initialized with high pass filter in SRM. Table 2 reported the MSE and SNR for both carrier and secret for two settings under TIMIT and LibriSpeech validation set. Note that, both the MSE and SNR are at a relatively normal level, which meant the proposed method could produce high fidelity secret audio and less distortion steganographic audio.
\begin{table}[]
	\centering
	\caption{MSE and SNR for both carrier loss and secret loss with two training settings under two datasets.}
	\setlength{\tabcolsep}{1.4mm}
	\begin{tabular}{cccccc}
		\hline
		& Setting&Carrier Loss& Secret Loss&Carrier SNR& Secret SNR \\
	    \hline
		TIMIT       
		& RAN  & 0.0001   & 0.0003   & 6.8479 & 0.3599     \\
		& HPF & 0.0001    & 0.0004   & 7.2413 &-2.4235     \\
		LibriSpeech 
		& RAN  & 0.0035   & 0.0047   & 1.8732 & 0.1239      \\
		& HPF & 0.0056    & 0.0058   & 2.2481 & 0.1272      \\
	\hline
	\end{tabular}
\end{table}
\begin{figure}
	\includegraphics[width=1.0\linewidth]{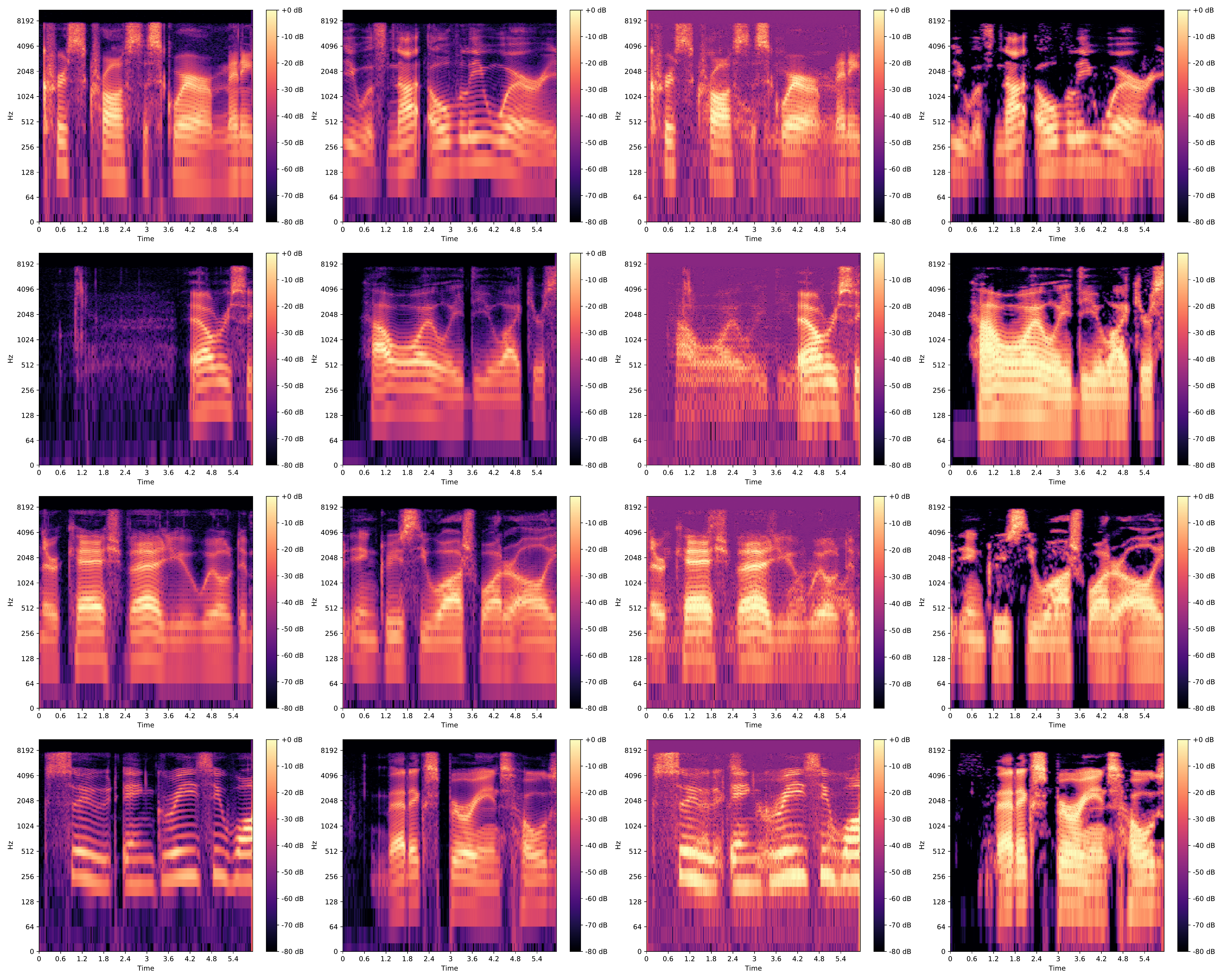}
	\centering
	\begin{center}
		\caption{Steganography and decoding spectrogram results. The horizontal axis represents the time while the vertical axis represents the frequency. The color of the figure represents the power level of the audio. The first row is the carrier audio spectrogram. The second row is the secret audio spectrogram. The third row is steganographic audio spectrogram. The last row is the decoding secret spectrogram.} \label{fig3}
	\end{center}
\end{figure}
\par
Figure. 3 presents visualizations of carrier and secret and their transformation after steganography and decode operation in the form of spectrogram. The horizontal axis represents time while the vertical axis represents the frequency. The color of the figure represents the power level of the audio. Note that, there is almost no difference in waveform shape between the carrier and steganographic audio. However, all the spectrogram become lighter after the steganagraphy operation which indicates that embedding operation chose areas on each frequency. Although the difference is visually, it is almost indistinguishable to human listener.
\subsection{Robust Experiments}
In order to evaluate the robustness of our models, we implemented robust experiments. To simulate the real public channel, We added a Gaussian noise of 60 dB to the steganographic audio and test the extraction effect of the decoder. We used two training settings: NOR which meant normal training used in Section 4.3; AN which meant add a Gaussian noise of 60 dB to the carrier audio during the training process.\par
\begin{table}[]
	\centering
	\caption{The simulation of real public channel MSE and SNR for both carrier loss and secret loss with two training settings under two datasets}
	\setlength{\tabcolsep}{2mm}
	\begin{tabular}{cccccc}
		\hline
		& settings	&   Carrier Loss& Secret Loss&Carrier SNR& Secret SNR \\
		\hline
		TIMIT    
		& NOR & 0.0001  & 0.0003  & 6.8470   & -0.0596    \\
		& AN & 0.0042   & 0.0002  & 5.9564   & 1.9642      \\
		LibriSpeech 
		& NOR & 0.0035  & 0.0050  & 1.8732   & -0.2861   \\
		& AN & 0.0096   & 0.0049  & 0.9148   & 0.1753     \\      
		\hline
	\end{tabular}
\end{table}
Table 3 reports the MSE and SNR between the carrier and steganographic audio under two training settings and two validation dataset. Note the second and fourth columns, when the noise interferes, the secret Loss MSE and secret SNR of NOR strategy increased and decreased respectively, which indicate that the steganography performance Weakened. The third and fifth columns report the results of AN training settings. The secret loss and secret SNR nearly back to the normal level, but the Carrier Loss and Carrier SNR decreased, which indicates the fidelity of steganographic audio decreased. In general, the AN training setting could enhance the decoding robust performance.\par
\subsection{Security Experiment}
To test the security of our model, we implemented security experiment. The security is the key point in steganography. During the training process in our GAN based steganography, the encoder was trained to fool the steganalyzer. But it does not indicate the proposed method could counter the steganalyzer, due to the training strategy in GAN which limits the effect of the discriminator. Therefore, we took the additional experiment to show the proposed method could counter independent steganalyzer.\par 
We created additional 5,000 steganographic audios with the proposed steganographic algorithm. The original carriers and the steganographic audios are merged to form a new dataset, then split in half, creating a training set and a validation set. We compared our scheme against steganographic algorithms HCM and ECCS. For each steganographic algorithm, we trained both SRM and CNN based steganalyzer on the training set, and then reported the accuracy of the steganalyzer on the validation set.  Table 4 shows the accuracy of three methods compares against the rich model(SRM) and a designed CNN-based steganalyzer. Note that the proposed method performs well against other steganographic methods.
\begin{table}[]
	\centering
	\caption{The accuracy of three methods compares against the rich model(SRM) and a designed CNN based steganalyzer.}
	\setlength{\tabcolsep}{3mm}
	\begin{tabular}{ccccc}
		\hline
		&\multicolumn{2}{c}{TIMIT} &  \multicolumn{2}{c}{LibriSpeech}               \\
		\hline
		Algorithms & SRM   & CNN-steganalyzer & SRM      & CNN-steganalyzer \\
		GAN-ste    & 0.8526     & 0.9156      & 0.7382   & 0.9258             \\
		HCM        & 0.8614     & 0.8962      & 0.6841   & 0.8917             \\
		EECS       & 0.7130     & 0.8451      & 0.7024   & 0.9145            \\
		\hline
	\end{tabular}
\end{table}
\section{Conclusion}
In this paper, we proposed a novel automatically generated audio steganography method. The proposed model consists of three components which creating the embedding, extracting and discriminating process. We take the external detector into consideration and use a steganalyzer as discriminator during the adversarial training. We demonstrated the effectiveness of our model by implementing three different experiments. In spectrogram results, the model embeds the secret on all the area of the figure may cause the secret be discovered. We would like to further investigate some adaptive method in future work. The encoder loss is formed by the sum of decoder loss and steganalyzer loss, finding the balance between three losses could be another research direction. We would like to consider other types of GAN such as CycleGAN to the steganography.
\section*{Acknowledgement}
This work was partially supported by the National Key Research Development Program of China (2016QY01W0200), the National Natural Science Foundation of China NSFC (U1636101, U1636219, U1736211).
% the environments 'definition', 'lemma', 'proposition', 'corollary',
% 'remark', and 'example' are defined in the LLNCS documentclass as well.

\end{document}